\documentstyle[12pt]{article}
\begin{document}
\rightline{July 1994}
\rightline{Revised, July 1994}
\rightline{McGill/94-36}
\vskip 2cm
\begin{center}
\begin{large}
Experimental signatures of a massive mirror photon
\vskip 2cm
\end{large}
R. Foot\\
Department of Physics,\\
McGill University, 3600 University Street,\\
Montreal, Quebec, Canada H3A 2T8.\\
\vskip 1cm

\end{center}
\vskip 1cm
ABSTRACT
\vskip .5cm
\noindent
There has been some speculation about the possible existence
of a second photon which only couples to ordinary matter
through mixing of the gauge kinetic terms.
If the exotic photon is massless, then the principle
phenomenological effect of the kinetic mixing is
to give the mirror particles small electric charges
(mini-charged particles).
We examine the possibility that the exotic photon is massive.
In this case the experimental signatures of the mirror photon
are quite different to the case when the exotic photon is massless.
In this scenario you do not get mini charged fermions
but instead you can get processes directly producing the mirror
photon such as Compton-like scattering: $e+\gamma \rightarrow e + \gamma^m$
(where $e, \gamma, \gamma^m$ denote the electron, photon and mirror photon
respectively).

\newpage

There has been some speculation in the past concerning
the possible existence of a second photon (which
we will denote as mirror photon) [1,2,3,4].
In particular models realizing parity as an exact
unbroken symmetry predict the existence of a second
photon [3]. There are also models with parity spontaneously
broken which predict a second photon [4].

In gauge models with two $U(1)$ gauge factors,
the two $U(1)$ gauge boson kinetic terms
can mix:
$${- \cal L}_{mix} = {\omega \over 4} F^1_{\mu \nu} F^{2 \mu \nu}, \eqno (1)$$
where $F^{1,2}_{\mu \nu} = \partial_{\mu} A^{1,2}_{\nu} -
\partial_{\nu} A^{1,2}_{\mu}$. Such a mixing term is
gauge invariant, renormalizable and allowed in the bare Lagrangian.
It does not occur in the standard model
because the standard model has only one $U(1)$ factor.

If the mirror (and ordinary) photons are massless, then
the principle effect of the kinetic mixing is to effectively
couple the mirror fermions
to ordinary photons, thus naturally leading to mini-charged
particles. This is an important phenomenological effect and
has been discussed previously [1-4].
Experimental bounds on minicharged particles are very weak (for
a summary of the experimental and astrophysical bounds see Ref. [5] and
references there-in)

Another possibility is that the mirror photons are massive.
This can be achieved by assuming that the mirror electromagnetism
is spontaneously broken. (This can be achieved in, for example,
models where parity connects the ordinary and mirror worlds,
but is spontaneously broken in such a way as to break the
mirror electromagnetism). This is the possibility that we
will consider in this note.

The gauge boson kinetic terms (with mixing) have the form:
$${- \cal L}_{kin} = {1 \over 4} F_{1\mu\nu}F_1^{\mu\nu} + {1 \over 4}
F_{2\mu\nu}F_2^{\mu\nu} + {\omega \over 4}F_{1\mu\nu}F_2^{\mu\nu},
\eqno (2)$$
where $A_1$ resembles the ordinary photon and couples only
to ordinary matter, while $A_2$ is a mirror photon which
only couples to mirror matter. We will assume for simplicity
that the mirror $U(1)$
has the same coupling ($e$) as ordinary electromagnetism (which
can be achieved in models with a parity symmetry connecting the
ordinary and mirror worlds).

The kinetic terms can be put into canonical form:
$${- \cal L}_{kin} = {1 \over 4} F{'}_{1\mu\nu}F{'}_1^{\mu\nu}
+ {1 \over 4} F{'}_{2\mu\nu} F{'}_2^{\mu\nu}, \eqno (3)$$
by the transformation:
$$A'_1 = \alpha A_1 + \beta A_2$$
$$A'_2 = \gamma A_1 + \delta A_2  \eqno (4)$$
where $\alpha^2 + \gamma^2 = \beta^2 + \delta^2 = 1$ and
$2\alpha \beta + 2 \gamma \delta = \omega$.

If the mirror photon is massive,
then the photon mass eigenstate basis is one with
$\gamma = 0$. Thus, the mass eigenstate basis is:
$$A'_1 = A_1 + {\omega \over 2} A_2$$
$$A'_2 = \sqrt{1 - ({\omega \over 2})^2}\ A_2 \eqno (5)$$
The state $A'_1$ is massless and is identified as the photon.
The state $A'_2$ is massive and is the physical (i.e. mass eigenstate)
mirror photon. By inverting eq.(5) we see that
the photon, $A'_1$,  couples only to
ordinary matter, while the mirror photon, $A'_2$ couples both
to ordinary matter and to mirror matter. This situation is quite
different to the case when the mirror photon is massless. In
that case, the mirror photon does not couple directly to
ordinary matter [1,6]

An important parameter is the mass of the mirror photon ($M$).
If $M$ is larger than twice the mass of the electron,
then the mirror photon can decay into electron positron pairs [7].
This will be the dominate decay if $M$ is lighter than the mirror
fermions (but more massive then twice the mass of the electron).
The decay rate of the mirror photon into electron positron pairs
can easily be calculated (to leading order in $\omega$):
$$\Gamma (A'_2 \rightarrow e^+e^-) =
\left({\omega \over 2}\right)^2{\alpha M \over 3}, \eqno (6)$$
where $\alpha = e^2/4\pi$ and I have neglected the phase space factor.
This width corressponds to a lifetime of
$$ \tau_{A'_2} \approx  (5/2) \times 10^{-19}
\left({\omega \over 2}\right)^{-2} ({MeV \over M})
\ \ {\rm seconds}.  \eqno (7)$$
Thus, unless $\omega$ is very tiny, we expect the mirror photon
to decay rapidly if its mass is greater than about 1 MeV.

It is interesting to note that consistency with the standard
big bang model of cosmology can imply a {\it lower bound} on $\omega$.
If the mirror photon decays fast enough (within 1 second), then
the mirror photon does not contribute significantly
to the energy density of the universe during the nucleosynthesis
era [8]. This condition implies that
$$\left({\omega \over 2}\right) \sqrt{M \over MeV} \
>  5 \times 10^{-10}.
\eqno (8)$$

We argue that the best way to experimentally
search for the effects of the kinetic
mixing is through processes directly producing mirror photons.
In particular consider the production of mirror
photons ($\gamma^m$) via Compton-like scattering
$\gamma + e \rightarrow \gamma^m + e$. The
relative cross-section of this process is
$${\sigma(\gamma e \rightarrow \gamma^m e) \over
\sigma(\gamma e \rightarrow \gamma e)} = (\omega/2)^2, \eqno (9)$$
where I have neglected the phase space factor.
If the mirror photon has mass less than about 1 MeV, then it should
either be long lived, or decay rapidly into mirror matter (if there is
mirror matter lighter than the mirror photon). In either of these cases,
the Compton scattering process can provide a clear experimental signature
of the mirror photon since the mirror photon can pass undetected
through the detector. The existence of the mirror photon can be inferred due
to the missing energy and momentum.

If the mirror photon is more massive than about 1 MeV, then as discussed
earlier, it
can decay into electron positron pairs.
In this case it is expected to decay while in the detector.
This case should also provide a clear signature for the mirror photon.

Finally, the mirror photon could also appear in many other processes.
For example, in rare decays of the neutral pion.
Some of the time, the pion will decay into a ordinary ($\gamma$) and
mirror photon ($\gamma^m$) (provided that it is kinematically allowed,
i.e.  $M < m_{\pi}$). The relative decay rate for this process is:
$${\Gamma (\pi^0 \rightarrow \gamma \gamma^m) \over
\Gamma (\pi^0 \rightarrow \gamma \gamma)} = (\omega/2)^2. \eqno (10)$$

We believe that the above experimental tests should be able to
place the best bound on the kinetic mixing parameter $\omega$, or possibly
discover the kinetic mixing phenomenon. They also have the advantage
of being model independent. They only assume the existence of a second
massive photon, and nothing about the exotic fermion content.

\vskip 1cm
\noindent
{\bf
Acknowledgement:}
\vskip .39cm
\noindent
I would like to thank H. Lew and M. Peskin for important discussions,
which lead to major revisions of the paper.

\newpage
\vskip 1cm
\noindent
{\bf
References:}

\vskip 1cm
\noindent
[1] B. Holdom, Phys. Lett. B166, 196 (1986).
\vskip .5cm
\noindent
[2] S. L. Glashow, Phys.  Lett. B167, 35 (1986); E. D. Carlson
and S. L. Glashow, Phys. Lett. B193, 168 (1987).
\vskip .4cm
\noindent
[3] R. Foot, H. Lew, and R. R. Volkas, Phys. Lett. B272, 67 (1991);
Mod. Phys. Lett. A7, 2567 (1992); R. Foot, Mod. Phys. Lett. A9, 169 (1994).
\vskip .4cm
\noindent
[4] S. M. Barr, D. Chang, and G. Senjanovic, Phys. Rev. Lett. 67, 2765 (1991);
R. Foot and H. Lew (to appear).
\vskip .4cm
\noindent
[5] S. Davidson and M. Peskin, Phys. Rev. D49, 2114 (1994).
\vskip .4cm
\noindent
[6] I would like to thank H. Lew and M. Peskin for helping me
understand this point.
\vskip .4cm
\noindent
[7] If the mirror photon is less massive than twice the electron mass then
the dominate decay mode should be the 3 photon decay ($\gamma^m \rightarrow
3\gamma$). This decay arises from the Feynman 1-loop diagram (with
a virtual electron in the loop). Note that the 2 photon decay of the massive
mirror photon is forbidden by Yang's theorem.
\vskip .4cm
\noindent
[8] Note that if $\omega$ is very small then the mirror photon
decouples from ordinary matter. For
this reason it is possible to get around
the lower bound on $\omega$. However, in many specific models, there
are other interactions which bring the mirror and ordinary sectors
into thermal contact (such as scalar interactions).
The applicability of the lower bound eq.(8) is thus model dependent.
\end{document}